\documentclass[a4paper,11pt]{article}
\usepackage{pos}

\usepackage{physics}
\usepackage[utf8]{inputenc}
\usepackage[T1]{fontenc}    
\usepackage{slashed}
\usepackage{mathtools}        
\usepackage{bbm}              
\usepackage{bm}               
\usepackage{caption}
\usepackage{subcaption}
\usepackage{multirow}
\usepackage{array}
\usepackage{booktabs} 
\usepackage[most]{tcolorbox}

\usepackage{graphicx}
\usepackage{comment}
\usepackage{latexsym}
\usepackage{xspace}
\usepackage{lscape}
\usepackage{color}
\usepackage{hyperref}
\usepackage{relsize}
\usepackage{tabularx}
\usepackage{amsmath}
\usepackage{lipsum}
\usepackage{feynmp-auto}
\usepackage{upgreek}
\usepackage{nicefrac}
\usepackage[titletoc]{appendix}

\usepackage{cleveref} 
\crefname{equation}{}{}
\crefname{figure}{}{}    
\crefname{table}{}{}
\crefname{section}{}{}   
\crefname{appendix}{}{}
\crefname{footnote}{}{}

\crefalias{subequation}{equation}

\def\nc{\newcommand}

\newcommand{\be}{\begin{equation}}
\newcommand{\ee}{\end{equation}}
\newcommand{\bea}{\begin{eqnarray}}
\newcommand{\eea}{\end{eqnarray}}
\newcommand{\ba}{\begin{array}}
\newcommand{\ea}{\end{array}}
\renewcommand{\vec}[1]{\bm{#1}}

\nc{\nn}{\nonumber}

\nc{\deldag}{{\mathbin{\partial\mkern-10mu/}}}
\nc{\kdag}{{\mathbin{k\mkern-10mu /}}}
\nc{\udag}{{\mathbin{u\mkern-10mu /}}}
\nc{\Ddag}{{\mathbin{D\mkern-10mu /}}}

\def\Slashnew#1{#1\kern-0.55em\raise.05ex\hbox{/}}
\def\slashnew#1{#1\kern-0.5em\raise.05ex\hbox{{$\scriptstyle /$}}}

\def\lsim{\mathrel{\raise.3ex\hbox{$<$\kern-.75em\lower1ex\hbox{$\sim$}}}}
\def\gsim{\mathrel{\raise.3ex\hbox{$>$\kern-.75em\lower1ex\hbox{$\sim$}}}}

\nc{\shalf}{\ensuremath{\textstyle \frac{1}{2}}}
\nc{\ihalf}{\ensuremath{\textstyle \frac{i}{2}}}

\def\sfrac#1#2{\ensuremath{\textstyle \frac{#1}{#2}}}

\def\emph#1{{\em #1}}

\def\eg{{\em e.g.}}

\nc{\sss}{\scriptscriptstyle}
\nc{\W}{{\sss W}}
\nc{\sparallel}{{\sss\parallel}}
\nc{\tGamma}{{\tilde \Gamma}}

\newcommand{\evec}[1]{{\bm{#1}}} 

\def\l{{\scriptscriptstyle <}}
\def\g{{\scriptscriptstyle >}}

\nc{\CPslash}{{{\scriptscriptstyle {\rm CP}}\mkern-18mu / \mkern 10mu}}

\def\W{{{\cal W}}}
\nc{\HF}{{\sss \rm FH}}

\title{Quantum kinetic equations with flavor and particle-antiparticle coherences for neutrinos}
\ShortTitle{QKE's for mixing neutrinos}

\author[a,b]{Kimmo Kainulainen}
\author*[a,b]{Harri Parkkinen}

\affiliation[a]{Department of Physics, PL 35 (YFL), 40014 University of Jyv\"askyl\"a, Finland}

\affiliation[b]{Helsinki Institute of Physics, PL 64, 00014 University of Helsinki, Finland}

\emailAdd{kimmo.kainulainen@jyu.fi}
\emailAdd{harri.h.parkkinen@jyu.fi}

\abstract{We develop a formalism to model neutrino evolution encompassing both flavor and particle-antiparticle mixings and decohering collisions. Our results include a quantum kinetic equation (a set of coupled scalar equations) for the generalized neutrino density matrix, valid for arbitrary neutrino masses and kinematics, and a comprehensive set of Feynman rules to compute collision integrals for coherently evolving states. We expose a novel shell structure describing the phase space of mixing neutrinos and show how the prior information on the system can enter into the theory and modify the neutrino flavor evolution. Potential applications of our results include modelling neutrino distributions in hot and dense environments and studies of neutrino mixing effects in colliders and in the early Universe.}

\ConferenceLogo{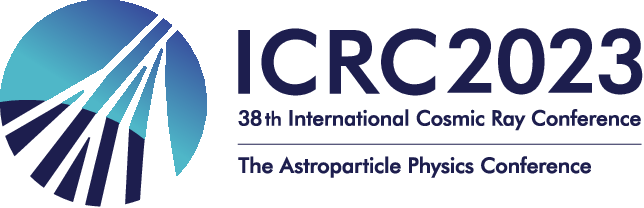}

\FullConference{%
38th International Cosmic Ray Conference (ICRC2023)\\
  26 July - 3 August, 2023\\
  Nagoya, Japan}

\begin{document}
\maketitle

\section{Introduction}
\label{sec:introduction}

\noindent Practically solvable quantum kinetic equations (QKE's) which can model accurately coherent neutrino evolution in presence of decohering collisions are necessary in many applications in neutrino physics~\cite{Volpe:2023met}. QKE's including the flavour coherences at different approximations have been known for some time~\cite{Enqvist:1991qj,Sigl:1992fn,McKellar:1992ja} and spatially homogeneous equations
including particle-antiparticle mixing can be found \eg~in~\cite{Her081,Her10,Fid11,Juk21}. Some aspects of particle-antiparticle mixing were considered also in~\cite{Serreau:2014cfa,Vlasenko:2013fja,Blaschke:2016xxt}. A fully self-consistent derivation of the QKE's, which include both the forward scattering potentials and the decohering collision integrals and encompass both flavour and particle-antiparticle mixing coherences has still been missing until now. Here we report a work~\cite{Kainulainen:2023ocv} that fills this gap. We clarify the role and distinction between the flavour and antiparticle oscillations and also derive simpler equations without the particle-antiparticle mixing. These equations are sufficient for description of the neutrino mixing and interactions in hot and dense environments and also for studying heavy neutrino oscillations in collider experiments and in different contexts in the early Universe, such as Leptogenesis or BBN. 

Our derivation is based on the full Schwinger-Dyson (SD) equations and the Closed Time Path formulation. In a few clearly justified steps we reduce the SD-equations to a set of scalar quantum kinetic equations, which include the information of flavor and particle-antiparticle mixing~\cite{Kainulainen:2023ocv}. Our work is based on earlier work in~\cite{Her081,Her10,Fid11,Juk21}. Our derivation assumes only adiabatically varying background fields, the validity of weak coupling expansion and eventually the spectral limit. As a result our equations are valid for arbitrary neutrino masses and kinematics. An integral part of the derivation is the introduction of a projective representation which reduces the SD-equation into a set of Boltzmann-type transport equations which contain all information of flavor or particle-antiparticle mixing. It also directly exposes a novel shell structure in the weak coupling limit: in addition to the usual mass shells new "coherence shells" emerge, that carry information about the particle-antiparticle coherences. In the UR-limit our equations become diagonal in the particle-antiparticle mixing to order $m/E$ and their flavour structure is greatly simplified. These equations are sufficient for most astrophysics and collider applications whereas the full equation is needed \eg~for problems involving particle production during preheating~\cite{Kainulainen:2021eki,Kainulainen:2022lzp}.

%
\section{Quantum kinetic equations}
\label{sec:quantum_kinetic_equations}
%
 
\noindent Coherently mixing out-of-equilibrium systems can be described by the Schwinger-Dyson equation which is equivalent to a coupled set of Kadanoff-Baym (KB) equations for real-time valued correlation functions. KB equations are manifestly non-local and the statistical functions are directly coupled to the pole functions.  To get a single local quantum kinetic equation (QKE), the pole equations must be decoupled from the statistical ones and the latter must be localized. In the Wigner space the localization translates to a truncation of the infinite order gradient expansion. This can be justified by the assumption of adiabatic background fields, or enforcing it by integrating over the momentum variables. The decoupling problem can be handled by splitting the statistical function into a background part, which is strongly coupled to the pole functions, and to a perturbation, whose equation formally decouples (this formal decoupling allows a wide range of solutions which makes the decoupling exact). For details of the procedure see~\cite{Kainulainen:2023ocv}. The resulting decoupled QKE for the local neutrino Wightman function $\smash{{\bar S}_\evec{k}^<}(t,t)$ reads: 
\begin{equation}
    \partial_t {\bar S}_\evec{k}^\l 
    + \frac{1}{2}\{\evec{\alpha}\cdot\nabla, \, {\bar S}_\evec{k}^\l\}
    = -i\big[{\cal H}_\evec{k} ,  \bar S^\l_\evec{k} \big]
     + i \Xi^\l_\evec{k} + \bar {\cal C}^\l_{{\rm H},\evec{k}},
    \label{eq:master}
\end{equation}
where $\smash{\alpha = \gamma^0 \gamma^i}$ and the Hamiltonian is $\smash{{\cal H}_\evec{k} = \bm{\alpha} \cdot \evec{k}\delta_{ij} + m_i\delta_{ij}\gamma^0}$. The forward scattering term $\smash{\Xi^\l_\evec{k}}$ and the Hermitian part of the collision term $\smash{\bar{\cal C}^\l_{{\rm H},\evec{k}}}$ are given in~\cite{Kainulainen:2023ocv}. Equation~\cref{eq:master} holds all information about coherence evolution for mixing neutrinos, but it is not yet useful for practical purposes.

To make the analysis and the interpretation of the results more convenient we write Eq.~\cref{eq:master} in the {\em projective representation}, constructed utilizing the helicity and vacuum Hamiltonian eigenbases. Indeed, without loss of generality we can parametrize the Wightman functions in adiabatic systems as follows:
\begin{equation}
\bar S_{\evec{k} ij}^\l(t,\evec{x}) = \sum_{h a a'}  f_{\evec{k} h ij}^{\l aa'}(t,\evec{x}) 
P_{\evec{k} h i j}^{a a'},
\label{eq:correlator_par}
\end{equation}
where $\smash{f_{\evec{k} h ij}^{\l a a'}(t,\evec{x})}$ are some unknown distribution functions (or the density matrix elements) and we defined the projection operator:
\begin{equation}
P_{\evec{k} h i j}^{a b} = N_{\evec{k}ij}^{a b} P_{\evec{k} h} P_{\evec{k} i}^a \gamma^0 P_{\evec{k} j}^{b},
\end{equation}
where the helicity and the vacuum energy projection operators read
\begin{equation}
P_{\evec{k} h} \equiv \frac{1}{2} 
   \big( \mathbbm{1} + h \bm{\alpha} \cdot \hat{\evec{k}} \gamma^5 \big )
\qquad \textrm{and} \qquad 
P_{\evec{k} i}^{a} \equiv \frac{1}{2} 
   \big( \mathbbm{1} + a \frac{\mathcal{H}_{\evec{k} i}}{\omega_{\evec{k} i}} \big ).
\end{equation}
Here $\smash{h= \pm1}$ is the helicity, $\smash{a,b=\pm1}$ are the energy sign indices, $i,j$ are the flavor indices and $\smash{\omega_{\evec{k} i}= (\evec{k}^2+m_i^2)^{1/2}}$ is the vacuum energy of the neutrino eigenstate. These projection operators satisfy completeness, orthogonality, and idempotence relations. 
The normalization factor is chosen as  $N_{\evec{k} ij}^{a b} \equiv \sqrt{2}(1 + ab (\gamma_{\bm{k}i}^{-1} \gamma_{\bm{k}j}^{-1} - v_{{\bm k}i}v_{{\bm k}j})^{-1/2}$, with $\gamma_{\bm {k}i}^{-1} = m_i/\omega_{\bm {k}i}$ and $v_{\bm{k}i} = |{\bm k}|/\omega_{\bm{k}i}$. With this choice the distribution functions will get the usual normalization in the thermal limit.

Utilizing the projective representation (\ref{eq:correlator_par}), multiplying with $\smash{P_{\vec{k} h j i}^{e'e}}$ and taking trace over the Dirac indices, it is simple task to reduce equation (\ref{eq:master}) to a set of scalar equations:\footnote{
%
%
We assumed that in the forward scattering terms in~\cref{eq:AH4} the background solution has the same form as the perturbation. This is strictly speaking true only in the spectral limit~\cite{Kainulainen:2023ocv}, which is what we are implicitly assuming here.}
%
%
\begin{equation}
    \boxed{
        \begin{aligned}
             \partial_t  f_{\vec{k} h ij}^{\l e e'} + ({\cal V}_{\bm{k} h ij}^{e'e})_{aa'} \hat{\bm k} \cdot \bm{\nabla}  f_{\vec{k} h ij}^{\l a a'} = & -2i\Delta\omega_{\bm{k} ij}^{e e'}  f_{\vec{k} h ij}^{\l e e'} 
            + \Tr[ \bar{\mathcal{C}}^\l_{{\rm H},\vec{k}h ij}   P_{\vec{k} h ji}^{e' e}] 
            \\
            & - i(\W^{{\rm H}ee'}_{\evec{k}hij})^{l}_{a} f_{\vec{k} h lj}^{\l a e'}
            + i[(\W^{{\rm H}e'e}_{\evec{k}hji})^{l}_{a}]^* f_{\vec{k} h il}^{\l e a},
        \label{eq:AH4}
        \end{aligned}
    }
\end{equation}
where $\hat{\bm k} = {\bm k}/|{\bm k}|$ and the repeated indices, $a$ and $l$, are summed over. Frequency sign indices $e,e'$ define the oscillation frequency:
\begin{equation}
2\Delta \omega_{\vec{k} ij} ^{ee'} \equiv \omega^e_{\vec{k}i} - \omega^{e'}_{\vec{k}j},
\label{eq:shell-frequencies}
\end{equation}
with $\omega^e_{{\bm k}i} \equiv e\omega_{{\bm k}i}$. The forward scattering tensor reads
\begin{equation}
(\W^{{\rm H}ee'}_{\evec{k}hij})^{l}_{a}  \equiv 
\Tr[P_{\vec{k} h ji}^{e'e} \bar{\Sigma}^{\rm H}_{\vec{k}il}(\omega^a_{\bm {k}i}) P_{\vec{k} hlj}^{a e'}],
\label{eq:W-tensor}
\end{equation}
and the velocity tensor can be written as
\begin{equation}
({\cal V}_{\vec{k} h ij}^{e'e})_{aa'} = \delta_{a'e'} {\cal V}_{\vec{k} h ij}^{eae'}
+ \delta_{ae} {\cal V}_{\vec{k} h ji}^{a'e'e},
\label{eq:isoD-def}
\end{equation}
where
\begin{equation} 
  {\cal V}_{\vec{k} h ij}^{a b c} \equiv \frac{1}{2} N_{\vec{k} ij}^{ac} N_{\vec{k} ij}^{bc}  
  \Big( v_{{\bm k}i} \Big[ \frac{a}{(N_{\vec{k} ij}^{bc})^2} 
  + \frac{b}{(N_{\vec{k} ij}^{ac})^2}\Big] - v_{{\bm k}j}c\delta_{a,-b} \Big).
  \label{eq:Dtensor}
\end{equation}
The collision term can be expressed in multiple different ways. An especially useful form is
\begin{equation}
    \Tr[ \bar{\mathcal{C}}^\l_{{\rm H},\vec{k}hij} P_{\vec{k} h ji}^{e' e}] 
    = \frac{1}{2} \Big( (\W^{{\g}ee'}_{\evec{k}hij})^l_a   f_{\vec{k} h lj}^{\l a e '}
                      + [(\W^{\g e'e}_{\evec{k}hji})^l_a]^*f_{\vec{k} h il}^{\l e a } 
                      - ( > \leftrightarrow < ) \Big),
\label{eq:collision_term}
\end{equation}
where the sum over $l$ and $a$ is again implied and the ${\cal W}^{s}$-tensors are defined similarly to~\cref{eq:W-tensor} with $\bar \Sigma^{\rm H} \rightarrow \bar \Sigma^{\rm s}$, where $s = >,<$.

All terms in our master equation~\cref{eq:AH4} have a simple interpretation: The first term on the right hand side coming from the Hamiltonian commutator term determines the relevant oscillation times scales for different solutions. The second term in the right hands side of~\cref{eq:AH4} is the collision term and the terms in the second row are forward scattering corrections. The left hand side displays a generalized Liouville term with the velocity tensor that determines the effect of different group velocities on the coherence evolution. The apparent complexity of~\cref{eq:AH4} reflects the generality of the equation, which is valid for arbitrary neutrino masses and kinematics and includes all information of flavor and particle-antiparticle mixings.

We wrote the master equation~\cref{eq:AH4} using frequency states rather than particle-antiparticle solutions, since this is notationally much simpler. The positive frequency solutions correspond naturally to particles, and the negative solutions with inverted 3-momenta correspond to antiparticles, according the following relation for distribution functions: 
$\bar f^{\l,\g}_{\evec{k}hij} = - f^{\g,\l --}_{(-\evec{k})hij}$. Here functions with the bar refer to antiparticles. Using this replacement rule one can always transform the results between frequency solutions and particle-antiparticle solutions when needed. 

\paragraph{UR-limit.} 

The master equation~\cref{eq:AH4} simplifies a lot in the ultra-relativistic (UR) limit. If we define a
diagonal velocity matrix $v_{{\bm k}ij} \equiv \delta_{ij}|\evec{k}|/\omega_{\evec{k}i}$, we can write equation~\cref{eq:AH4} in the compact and familiar form of a density matrix evolution equation:
\begin{equation}
\boxed{
\partial_t  f_{\vec{k} h}^{e} 
           + \sfrac{1}{2} \{ v_{\bm k}, \hat{\bm{k}} \cdot \bm{\nabla} f_{\vec{k} h}^{e} \} =
- i[H_{\evec{k}h}^e, f_{\vec{k} h}^{e}] + \bar{\mathcal{C}}_{\vec{k} h}^{e},\strut}
\label{eq:AH2_UR}
\end{equation}
where $\smash{(\bar{\mathcal{C}}_{\vec{k} h}^{e})_{ij} \equiv {\rm Tr}[ \bar{\mathcal{C}}^\l_{{\rm H},\vec{k}hij} P_{\vec{k} h ji}^{ee}]}$ is the frequency diagonal collision integral and $(H^e_{\evec{k}h})_{ij}$ is the effective matter Hamiltonian:
\begin{equation}
(H^e_{\evec{k}h})_{ij} = e \delta_{ij}\omega_{\evec{k}i} + (V^e_{\evec{k}h})_{ij},
\label{eq:matter_hamiltonian}
\end{equation}
where $(V^e_{\evec{k}h})_{ij}$ is the standard forward scattering potential. With light neutrinos and relatively small propagation distances, one can further set $v_{{\bm k}ij} \rightarrow \delta_{ij}$, in which case the spatial gradient term reduces to $\smash{ \sfrac{1}{2} \{ v_{\bm k}, \hat{\bm{k}} \cdot \bm{\nabla} f_{\bm{k} h}^{e} \} \rightarrow \hat{\evec{k}}\cdot \bm{\nabla}f_{\bm{k} h}^{e}}$. Equation~\cref{eq:AH2_UR} is frequency diagonal and no longer contains the particle-antiparticle coherences. It still describes the strong coupling between particle- and antiparticle sectors via the matter potential term, which leads to many interesting phenomena in the neutrino mixing in the early Universe and in compact objects~\cite{Volpe:2023met}.

%
\section{Collision terms}
\label{sec_collision_terms}
%

\noindent Computation of the full collision term with flavor and particle-antiparticle mixing for arbitrary neutrino masses and kinematics has remained unsolved until now. Using our formalism the collision integrals are easy to evaluate however. We find that they can always be divided into a dynamical matrix elements squared and phase space elements, giving rise to the familiar structure:
\begin{equation}
    \begin{split}
    \overline{\mathcal{C}}_{{\rm H},\evec{k} h ij}^{< e e'} =  \sum_{Y} \nolimits
         \frac{1}{2\bar\omega_{\evec{k}lj}^{aa'}} \int\dd{\rm{PS}_3} 
         \Big[
         \sfrac{1}{2}(\mathcal{M}^2)_{\evec{k}  h i j \{\evec{p}_i,Y\}}^{ee'} 
         \Lambda_{\evec{k} hj\{\evec{p}_i,Y \},x}
         +  (h.c.)^{e'e}_{ji} \Big].
    \label{eq:collZZ1}
    \end{split}
\end{equation}
Note the flipping of indices in the Hermitian conjugate term, in accordance with~\cref{eq:collision_term}. We collected all summed indices into curly brackets with $Y \equiv \{ {\rm X}_i, h', a, a', l\}$ and defined a shorthand notation $A_{{\rm X}_i} \equiv A_{h_il_il_i'}^{a_ia_i'}$. All particle distribution functions were combined into $\Lambda \equiv \Lambda^\g-\Lambda^<$ with
\begin{equation}
\label{eq:lambda factor}
\Lambda^{\l, \g}_{\evec{k} hj\{\evec{p}_i, Y \},x} = 
     f^{\l,\g}_{{\rm X}_1\evec{p}_1}(x) \,
     f^{\g,\l}_{{\rm X}_2\evec{p}_2}(x)\,
     f^{\l,\g}_{{\rm X}_3\evec{p}_3}(x)\,
     f^{\g,\l a a'}_{\evec{k} h' l j}(x),
\end{equation}
and the phase space factor reads
\begin{equation}
    \label{eq:phase space standard}
    \int  \dd{\rm{PS}_3} \equiv \hspace{-0.5ex}\int 
      \hspace{-0.5ex}\Big[ \prod_{i=1,3} \frac{{\rm d}^3\evec{p}_i}{(2\pi)^3 2\bar\omega_{\evec{p}_il_il_i'}} \Big] 
        (2\pi)^4 \delta^4(k^a_l + p_{2 \, l_2}^{a_2} 
        - p_{1 \, l_1'}^{a_1'}-p_{3 \, l_3'}^{a_3'}).
\end{equation}

The matrix element squared $\smash{(\mathcal{M}^2)_{\evec{k} h ij \{\evec{p}_i Y\}}^{ee'}}$ contains all dynamical details related to the interaction process. It can be evaluated using a simple set of Feynman rules given in figure~\cref{fig:feynman_rules2}, where we used the $D$-tensor notation: 
$\smash{D_{\evec{k} h i j}^{ab} 
\equiv  2\bar\omega^{ab}_{\evec{k}ij} P^{ab}_{\evec{k}hij}\gamma^0 
\equiv  ab \hat N^{ab}_{\evec{k}ij}  P_{\evec{k}h}(\slashed k_{i}^a+m_i) (\slashed k_{j}^b+m_j)}$,
where we defined $(\smash{k_{i}^a)^\mu \equiv (a\omega_{\evec{k}i},\evec{k})}$ as well as $\smash{\,\hat N_{\evec{k}ij}^{ab} \equiv N_{\evec{k}ij}^{ab}\bar\omega^{ab}_{\evec{k}ij}/(2\omega^a_{\evec{k}i} \omega^b_{\evec{k}j})}$ and finally
$\smash{2\bar\omega^{ab}_{\evec{k}ij} \equiv a\omega_{\evec{k}i} + b\omega_{\evec{k}j} }$. 
These Feynman rules are to be used with the following instructions:

%
\begin{figure}[t]
\centering
\includegraphics[width=0.9\textwidth]{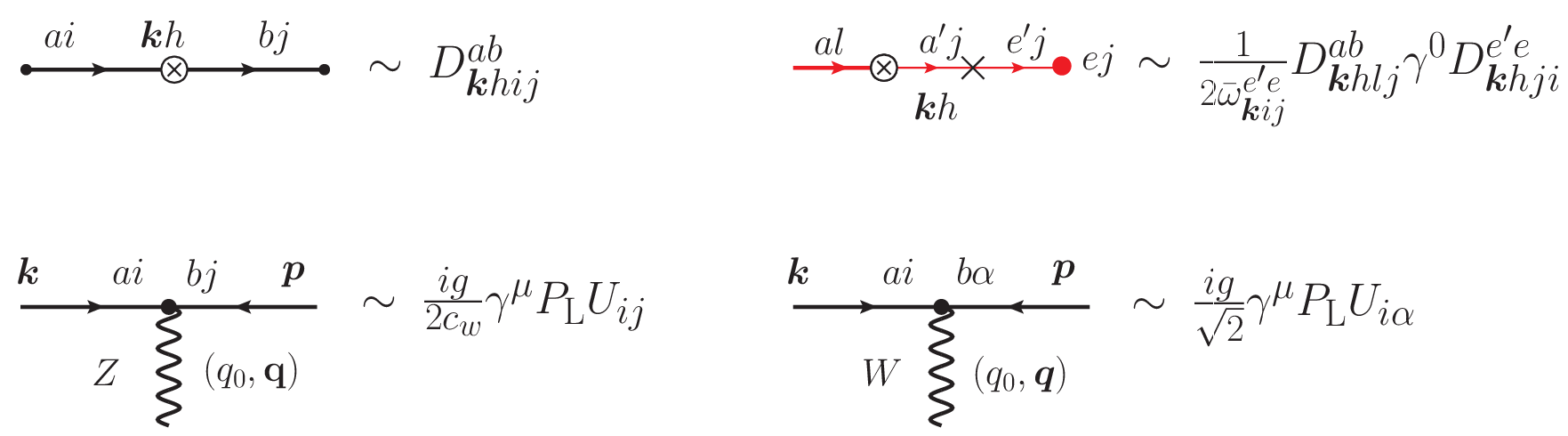}
\caption{Feynman rules for computing the squared matrix element for coherent neutrino states. The first propagator should be used for all internal lines and the red propagator (DMP) is used for the outgoing line in the diagram. In the $W$-boson vertex the matrix $U_{i\alpha}$ reduces to the usual PMNS-matrix in the case of pure active-active mixing (here $\alpha$ is lepton flavour). Similarly, in the pure active-active mixing  case, the matrix $U_{ij}$ in the $Z$-boson vertex reduces to $\delta_{ij}$. Finally $c_w = \cos\theta_w$. Figure is taken from~\cite{Kainulainen:2023ocv}.}
\label{fig:feynman_rules2}
\end{figure} 
%

\begin{figure}[t]
\centering
    \includegraphics[width=1.0\textwidth]{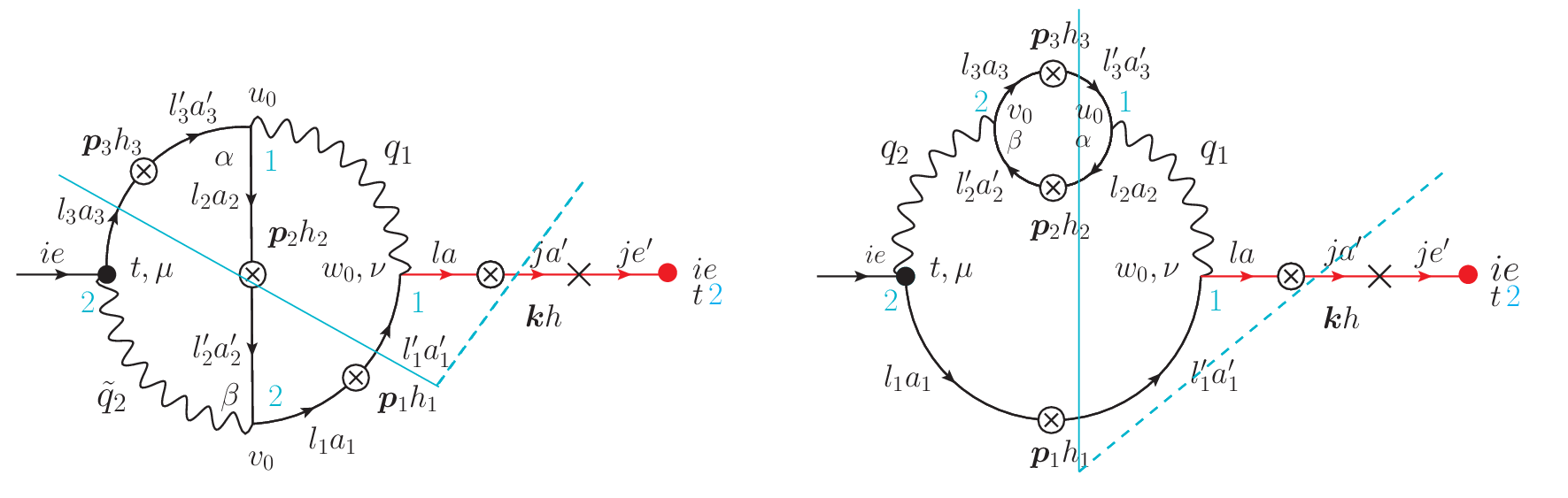}
\caption{Two-loop graphs contributing to neutrino-neutrino scattering with the explicit index structures and cuts. The right (direct, non-2PI) diagram contains the $s$- and $t$-channel processes and the left (interference, 2PI) diagram contains their interference. The black dot implies the starting point of the evaluation of the matrix element squared, and the red propagator is the DMP. The figure is taken from~\cite{Kainulainen:2023ocv}.}
\label{fig:ZZ combined}
\end{figure}

\begin{itemize}
\item{} Draw the loop diagrams that contribute to a given interaction process to the desired order in perturbation theory, and assign a unique momentum variable and flavor and frequency indices for each internal propagator line in the graph, allowed by the interaction vertices.
\item{} Assign the Keldysh-path indices to all vertices to isolate cuts that give rise to the desired interaction processes. You only need to evaluate $\Sigma^\g= \Sigma^{21}$ directly, so the first index is always 2 and the last 1. 

\item{} Read off the phase space functions contributing to the $\Lambda$-factor from all internal cut propagator lines. Add the phase space factor $f_{\evec{k} hlj}^{\l aa'}/2\omega^{aa'}_{\evec{k}hlj}$, associated with the external, dependent momentum propagator (DMP), marked red in diagrams in figure~\cref{fig:ZZ combined}.

\item{} Deduce the phase space density factor with the overall energy conserving delta function. This depends on the number of loops in the diagram and the cut one is interested in. 
\item{} Compute the matrix element squared using the Feynman rules shown in fig.~\cref{fig:feynman_rules2}. Start from the equivalent of the black dot shown in the diagrams in figure~\cref{fig:ZZ combined} and follow the direction of momentum in the graph. For each internal cut-line insert the standard propagator shown in the first diagram in~\cref{fig:feynman_rules2}. For each ("22") "11" line use the (anti) Feynman propagator. Add the DMP at the end of the fermion line it is connected to. Take a trace over the Dirac indices.
\item{} Divide the result by two and add the Hermitian conjugate accounting for the flip of indices as indicated in~\cref{eq:collZZ1}.
\end{itemize}
%
%
%

\paragraph{Neutrino-neutrino scattering.} 

As a demonstration, we give the squared matrix element for neutrino-neutrino scattering proceeding via $s$- and $t$-channels and their interference. This process corresponds to the Feynman diagrams and the cuts shown in~\cref{fig:ZZ combined}. Using the rules given above, one can immediately write the matrix elements squared for these processes:
\begin{equation}
(\mathcal{M}^2)_{\evec{k} h i j \{\evec{p}_i,{\rm Y}\}}^{ee'} \equiv
      \frac{1}{2\bar\omega_{\evec{k}ij}^{ee'}}
         \Tr[(A^{{\rm int},a}_{\evec{k}il\{\evec{p_i} {\rm X_i}\}} 
            + A^{{\rm dir},a}_{\evec{k}il\{\evec{p_i} {\rm X_i}\}})
         D_{\evec{k} h' l j}^{aa'} \gamma^0 D_{\evec{k} h j i}^{e'e}],
\label{eq:effective_matrix_element}
\end{equation}
where the leftmost 2-particle irreducible (2PI) interference diagram contributes the term
\begin{align}
A^{{\rm int},a}_{\evec{k}il\{\evec{p}_i,{\rm X}_i\}} = 
\Big(\frac{ig}{2c_w}\Big)^4 & \; U_{il_3} U_{l_3'l_2} U_{l_2'l_2} U_{l_2'l}
       {\cal D}_{Z\alpha\nu}(q_1) {\cal D}^{*}_{Z\mu\beta}(\tilde q_2) 
\nonumber \\ \times &
        \gamma^{\mu}    P_{\rm L} \, D_{{\rm X}_1\evec{p_1}}
        \gamma^{\alpha}    P_{\rm L} \, D_{{\rm X}_2\evec{p_2}}
        \gamma^{\beta} P_{\rm L} \, D_{{\rm X}_3\evec{p_3}}
        \gamma^{\nu}  P_{\rm L},
\label{eq:A_int}
\end{align}
and the rightmost non-2PI direct diagram gives
\begin{align}
A^{{\rm dir},a}_{\evec{k}il \{\evec{p}_i,{\rm X}_i\}} = 
-\Big(\frac{ig}{2c_w}\Big)^4 & \; U_{il_1} U_{l_1'l} U_{l_2'l_3} U_{l_3'l_2}
       {\cal D}_{Z\alpha\nu}(q_1) {\cal D}^{*}_{Z\mu\beta}(q_2) 
\nonumber \\ \times &
        \, \gamma^{\mu}     P_{\rm L} \, D_{{\rm X}_1\evec{p_1}}
        \gamma^{\nu}  P_{\rm L} \, 
        \Tr\big[\gamma^{\alpha} P_{\rm L} D_{{\rm X}_2\evec{p_2}} \gamma^{\beta}    P_{\rm L} 
                  D_{{\rm X}_3\evec{p_3}}\big].
\label{eq:A_direct}
\end{align}
Here the gauge boson momenta are $\smash{q_1 = (\omega^a_{\evec{k}l}\! - \omega^{a_1'}_{\evec{p}_1l_1'}; \evec{k} \!-\evec{p}_1)}$, $\smash{q_2 = (\omega^{a_3}_{\evec{p}_3l_3}\! - \omega^{a_2'}_{\evec{p}_2l_2'};\evec{k} \!-\evec{p}_1)}$ and $\tilde q_2 = (\omega^{a_1}_{\evec{p}_1l_1}\! - \omega^{a_2'}_{\evec{p}_2l_2'};\evec{k} \!-\evec{p}_3)$. The energy sign indices determine to which process (either a 2-2 scattering or a 1-3 decay) each term contributes. Here we gave the most general matrix element structure, which is however easy to evaluate using algebraic manipulation programs. The result simplifies dramatically in the UR-limit however, and becomes easily computable by hand~\cite{Kainulainen:2023ocv}.

%
\section{Weight functions}
\label{sec:weight_functions}
%

An essential feature of our proof, largely left aside in this brief note, was the localization of the KB-equations, which corresponded to an integration over the frequencies in the Wigner space~\cite{Kainulainen:2023ocv}. The physical reasoning for the integration is that one can never have a complete information about a given system. What matters from the point of view of the oscillation phenomenon, is the in general poor knowledge of the frequency and/or the momentum of the state in comparison to the phase space separation of the shell-solutions associated with the mixing. It is precisely this lack of information that allows for the oscillating solutions to exist. Indeed, if one {\em had} a precise information of the 4-momentum (\eg~due to very precise measurement of neutrino beam parameters), the separate mass-eigenstates would be emitted in the process, with no oscillation pattern left to study. 

This reasoning implies that the physical quantities that one {\em can} study are some coarse grained quantities which must carry the information about the preparation of the system into the theory describing their evolution. We can express this quantitatively by stating that the observable correlation function, that can be related to physically measurable quantities $\bar{k}$ and $\bar{x}$, is some weighted average of the original correlation function:
\begin{equation}
    S_{ij} (\bar{k}, \bar{x}) \equiv \frac{1}{(2 \pi)^4} \int {\rm d}^4x {\rm d}^4k  \;
    \mathcal{W}(\bar{k}, \bar{x}; k, x) S_{ij}(k, x),
    \label{eq:weighted average}
\end{equation}
where $\mathcal{W}$ is some weight function encoding the observationally accessible information about the system. Weight functions can affect any or all variables relevant for the problem. The parameters relevant for this paper were helicity frequency, 3-momentum, and spatial and temporal coordinates. For most problems the simple weight function we used (flat weight over the frequencies) and the ensuing master equations is sufficient. More general weight functions could be interesting avenue to derive adjustable and quantitative ways to take the effects of neutrino production and detection processes into account when describing neutrino evolution.

%
\section*{Acknowledgements}
%

\noindent
The work of HP was supported by grants from the Jenny and Antti Wihuri Foundation.


\bibliographystyle{ieeetr}
\bibliography{references.bib}

\begin{thebibliography}{10}

\bibitem{Volpe:2023met}
M.~C. Volpe, ``{Neutrinos from dense: flavor mechanisms, theoretical approaches, observations, new directions},''
\newblock arXiv:[\href{https://arxiv.org/abs/2301.11814} {2301.11814}], Jan 2023.

\bibitem{Enqvist:1991qj}
K.~Enqvist, K.~Kainulainen, and M.~J. Thomson, ``{Stringent cosmological bounds on inert neutrino mixing},'' {\em Nucl. Phys. B}, vol.~373, pp.~498--528, 1992.

\bibitem{Sigl:1992fn}
G.~Sigl and G.~Raffelt, ``{General kinetic description of relativistic mixed neutrinos},'' {\em Nucl. Phys. B}, vol.~406, pp.~423--451, 1993.

\bibitem{McKellar:1992ja}
B.~H.~J. McKellar and M.~J. Thomson, ``{Oscillating doublet neutrinos in the early universe},'' {\em Phys. Rev. D}, vol.~49, pp.~2710--2728, 1994.

\bibitem{Her081}
M.~Herranen, K.~Kainulainen, and P.~M. Rahkila, ``{Quantum kinetic theory for fermions in temporally varying backgrounds},'' {\em JHEP}, vol.~09, p.~032, 2008.

\bibitem{Her10}
M.~Herranen, K.~Kainulainen, and P.~M. Rahkila, ``{Coherent quantum Boltzmann equations from cQPA},'' {\em JHEP}, vol.~12, p.~072, 2010.

\bibitem{Fid11}
C.~Fidler, M.~Herranen, K.~Kainulainen, and P.~M. Rahkila, ``{Flavoured quantum Boltzmann equations from cQPA},'' {\em JHEP}, vol.~02, p.~065, 2012.

\bibitem{Juk21}
H.~Jukkala, K.~Kainulainen, and P.~M. Rahkila, ``{Flavour mixing transport theory and resonant leptogenesis},'' {\em JHEP}, vol.~09, p.~119, 2021.

\bibitem{Serreau:2014cfa}
J.~Serreau and C.~Volpe, ``{Neutrino-antineutrino correlations in dense anisotropic media},'' {\em Phys. Rev. D}, vol.~90, no.~12, p.~125040, 2014.

\bibitem{Vlasenko:2013fja}
A.~Vlasenko, G.~M. Fuller, and V.~Cirigliano, ``{Neutrino Quantum Kinetics},'' {\em Phys. Rev. D}, vol.~89, no.~10, p.~105004, 2014.

\bibitem{Blaschke:2016xxt}
D.~N. Blaschke and V.~Cirigliano, ``{Neutrino Quantum Kinetic Equations: The Collision Term},'' {\em Phys. Rev. D}, vol.~94, no.~3, p.~033009, 2016.

\bibitem{Kainulainen:2023ocv}
K.~Kainulainen and H.~Parkkinen, ``{Quantum transport theory for neutrinos with flavor and particle-antiparticle mixing},''
\newblock arXiv:[\href{https://arxiv.org/abs/2309.00881} {2309.00881}], Sep 2023.

\bibitem{Kainulainen:2021eki}
K.~Kainulainen and O.~Koskivaara, ``{Non-equilibrium dynamics of a scalar field with quantum backreaction},'' {\em JHEP}, vol.~12, p.~190, 2021.

\bibitem{Kainulainen:2022lzp}
K.~Kainulainen, O.~Koskivaara, and S.~Nurmi, ``{Tachyonic production of dark relics: a non-perturbative quantum study},'' {\em JHEP}, vol.~04, p.~043, 2023.

\end{thebibliography}

%
%
%

\end{document}